\documentclass[prl,showpacs,twocolumn,amsmath,amssymb]{revtex4}

\usepackage{psfrag}
\usepackage{graphicx}
\usepackage{color}

\usepackage[latin1]{inputenc}
\usepackage{graphicx}
\usepackage{amsmath}
\usepackage{amssymb}
\usepackage{relsize}

\newcommand{\pd}[2]{\frac{\partial #1}{\partial #2}}
\newcommand{\mean}[1]{\left \langle #1 \right \rangle}

\newcommand{\lam}{\lambda}
\newcommand{\beq}{\begin{equation}}
\newcommand{\eeq}{\end{equation}}

\newcommand{\nn} {\nonumber}

\newcommand{\xm} {u}
\newcommand{\xvar} {w}
\newcommand{\lf} {(\lam_f / \lam_i)}

\newcommand{\Wmin}{W^*}
\newcommand{\Wmina}{W^*}
\newcommand{\lmin}{\lambda^*}
 
\newcommand{\dxp} {d [x(\tau)]}
\newcommand{\pp} {p[x(\tau)]}

\newcommand{\qs} {qs}

\begin{document}

	\title{Optimal finite-time processes in stochastic thermodynamics}
\author{Tim Schmiedl and Udo Seifert}
\affiliation{{II.} Institut f\"ur Theoretische Physik, Universit\"at Stuttgart,
  70550 Stuttgart, Germany}

\begin{abstract}
{For a small system like a colloidal particle or a single biomolecule
embedded in a heat bath, the optimal protocol of an external control parameter minimizes the mean work required to drive the system from
one given
 equilibrium state to another  
in a finite time. In general, this optimal protocol obeys an
  integro-differential
equation. Explicite solutions both for
a moving laser trap and a time-dependent strength of such a trap show
finite jumps of the optimal protocol to be typical both at the beginning and the end of the
process.}
\end{abstract}


\def\ll{\lambda}
\def\la{\lambda(\tau)}
\def\ls{\lambda^*(\tau)}
\def\li{\lambda_i}
\def\wmus{W[\la]}

\maketitle

{\sl Introduction.--}
The concepts of classical thermodynamics like
applied work and exchanged heat can be applied to soft and biomatter systems as the study of the driven dynamics of single colloidal particles and of single 
biomolecules has shown \cite{bust05}.
Since thermal fluctuations contribute substantially, work and heat
acquire a stochastic contribution and must be described by
probability distributions. Exact results constraining such
distributions like the Jarzynski relation \cite{jarz97} and various generalizations thereof have 
been derived theoretically \cite{croo00, hata01, seif05a,spec05a, impa05} 
and tested experimentally \cite{wang02, carb04, trep04, schu05, blic06}. Typically, these exact relations hold for any
time-dependent driving described by an external control parameter $\la$.

In this paper, we ask for the optimal protocol $\ls$ that minimizes the
mean work required to drive such a system from one equilibrium state to
another in a {\sl  finite} time $t$. The emphasis on a finite time is
crucial since for infinite time the work spent in any quasi-static process
is equal to the free energy difference of the two states. For finite time
the mean work is larger and will depend on the protocol $\la$. 
Knowing the optimal protocol $\ls$ could inter alia improve the
extraction of free energy differences from finite-time path sampling
both in various numerical schemes \cite{hend01, humm01a, shir03,  ytre04, atil04, park04, lech06, ytre06} and in experimental studies \cite{liph02, coll05}.
Quite generally, the smaller the mean work is, the better 
the statistics for free energy estimates becomes \cite{atil04, jarz06}. A priori, one might expect
the optimal protocol connecting the given initial and final values to be
smooth as it was found recently in a case study within the linear response regime \cite{koni05}. 
In contrast, as our main result,  we find here  for genuine finite-time
driving that the optimal protocol involves discontinuities both at the
beginning and the end of the process.

For macroscopic systems,
optimal processes have been
investigated under the label of finite-time thermodynamics for quite some
time \cite{curz75, band82, andr84}. Indeed, jumps were found
there as well \cite{band82} despite the significant differences both in the role of heat
baths
and the equations of motion between macroscopic and stochastic thermodynamics. For the former, heat reservoirs 
of different temperature are typically involved in a search for optimal adaptions 
of, e.g.,  Carnot-like machines to finite time cycles.
In our context, the system always remains in contact with a single heat bath
of constant temperature $T$. Moreover, this heat bath provides thermal
fluctuations which require a stochastic formulation in contrast to the
deterministic description in macroscopic finite-time thermodynamics.

{\sl The model.--}
Paradigmatically, a Langevin equation describes the driven overdamped motion of a
single degree of freedom with co-ordinate $x$ in a time-dependent
one-dimensional potential $V(x,\la)$ as 
\beq \dot x = -\mu
\pd {V(x,\ll)} x + \zeta .  
\eeq 
Here,  $\mu$ is the mobility and time-derivatives are denoted by a dot
throughout the
paper.   The thermal
fluctuations are modelled as Gaussian white noise 
$ \mean{\zeta(\tau)
  \zeta(\tau')} = 2 \mu k_B T \delta(\tau-\tau'),  
$ with $k_B$ as Boltzmann's constant.
The time evolution of the probability distribution $p(x,\tau)$ to 
observe the particle at position $x$ at time $\tau$ is then governed by the Fokker-Planck equation
\beq
\partial_\tau p(x,\tau) = \partial_x \left[\mu \pd V x + \mu k_B T \partial_x \right] p(x,\tau).
\label{FP}
\eeq

Initially, the system
is in thermal equilibrium in the potential $V(x,\ll_i)$. During the
time-interval $0\leq \tau\leq t$, the control parameter $\la$ is varied from
$\ll_i$ to the final value $\ll_f$. The mean work spent in this process 
\beq
W[\la]=\int_0^t d\tau \dot\ll \mean{ \pd V \ll(x(\tau),\la) }
\label{meanw}
\eeq 
becomes a functional
of the protocol $[\la]$ where the average $\langle .... \rangle$ is over the
initial thermal distribution and over the noise history.
For notational simplicity, we set $k_BT=\mu =1$ in the following 
by choosing natural units for energies and times.
We will first investigate two case studies motivated by previously set up
experiments on colloidal particles and then analyze the  general case.

{\sl Case study I: Moving laser trap.--}
As an almost trivial, but still instructive 
introductory example, we consider a colloidal particle dragged through a viscous fluid by an optical tweezer with harmonic potential
\beq
V(x,\tau) = \left(x-\lam(\tau)\right)^2 / 2.
\eeq
The focus of the optical tweezer is moved according to  a protocol
$\lam(\tau)$. In previous experiments, such protocols have been used to test
the fluctuation theorem \cite{wang02} and the Hatano-Sasa relation \cite{trep04}. 
The optimal protocol $\ll^*(\tau)$ connecting given boundary values
 $\ll_i=0$ and
$\ll_f$ in a time $t$ minimizes the mean total work  (\ref{meanw}) which we express as a
functional of
the mean position of the particle  $\xm(\tau) \equiv \mean{x(\tau)}$
as
\begin{eqnarray}
\wmus &=& \int_0^{t} d\tau \dot \lam (\lam - \xm)  = \int_0^{t} d\tau  (\dot \xm + \ddot \xm) \dot \xm \nn \\
&=& \int_0^t d\tau \dot u ^2 + \left[\dot u ^2 \right]_0^t /2 . 
\label{W_mov}
\end{eqnarray}
Here, we have used 
\beq
\dot \xm = (\lam - \xm)
\label{dmean} 
\eeq which follows from averaging the Langevin equation.

The  Euler-Lagrange equation corresponding to (\ref{W_mov}),
$
\ddot \xm = 0
$, is solved by $u(\tau) = m\tau$, where $u(0)=0$ is enforced by the initial
condition. Eq. (\ref{dmean}) then requires the boundary conditions $\dot u(0) = \lam_i - u(0) = 0$ and $\dot u(t) = \lam_f - mt$ which can only be met by discontinuities in $\dot u$ at the boundaries which correspond to jumps in $\lambda$. Note that these ``kinks'' do not contribute to the integral in the second line of eq. (\ref{W_mov}). The yet unknown parameter $m$ follows from minimizing the
mean total work 
\beq
W =  m^2 t + (\lam_f - mt)^2 / 2
\eeq
which yields $m^{*} = \lam_f / (t+2)$. The minimal mean work $\Wmin =
\lam_f^2/(t+2)$ vanishes in the quasi-static limit $t \to \infty$. The optimal protocol then follows from eq. (\ref{dmean}) as
\beq
\ls = {\lam_f} (\tau + 1)/(t+2),
\eeq
for $0<\tau<t$.
As a surprising result, this optimal protocol implies two distinct symmetrical
jumps of size 
\beq
\Delta \ll \equiv\ll(0^+)-\ll_i=\ll_f-\ll(t^-)=\ll_f/(t+2)
\eeq
at the beginning and the end of the process.

A priori, one might have expected a continuous linear protocol 
$\ll^{\rm lin}(\tau)=\ll_f \tau/t$ to yield the lowest work but the explicite calculation
shows that 
\beq
W^{\rm{lin}} = (\lam_f / t)^2 (t + e^{-t} - 1) > \Wmin
\eeq
 for any
$t>0$, with a maximal value $W^{\rm{lin}}/W^*\simeq 1.14$ at $t\simeq 2.69$.

In macroscopic finite-time thermodynamics, the occurrence of such jumps has previously
been rationalized by pointing out the special nature of this type of 
variational problem where the highest derivative (here $\ddot u$ in eq. (\ref{W_mov})) occurs
linearly \cite{band82}. In the present model where fluctuations are
irrelevant to the mean work, these jumps have the same formal origin.

\begin{figure*}
 \includegraphics[width = 0.45 \textwidth]{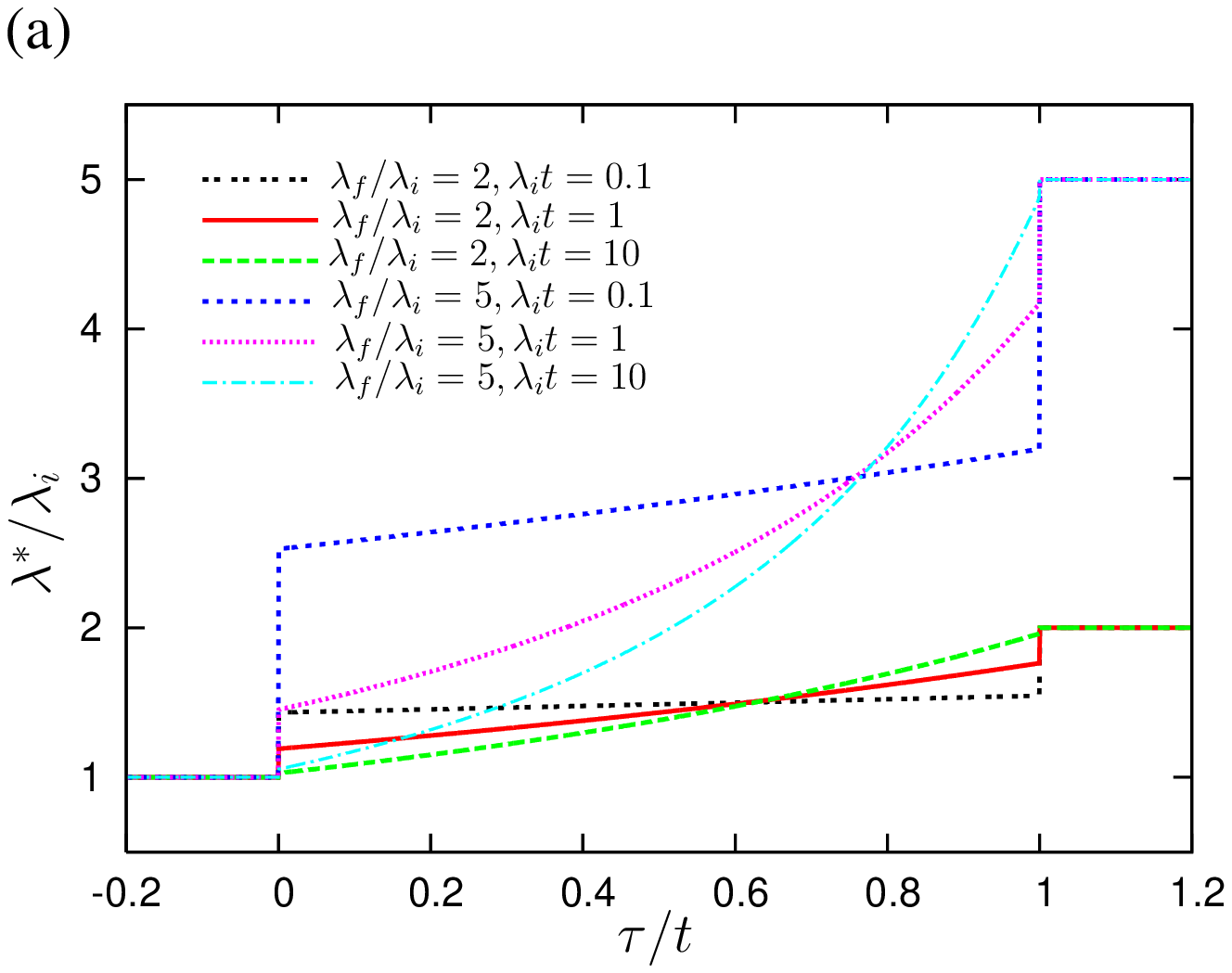}
\includegraphics[width = 0.44 \textwidth]{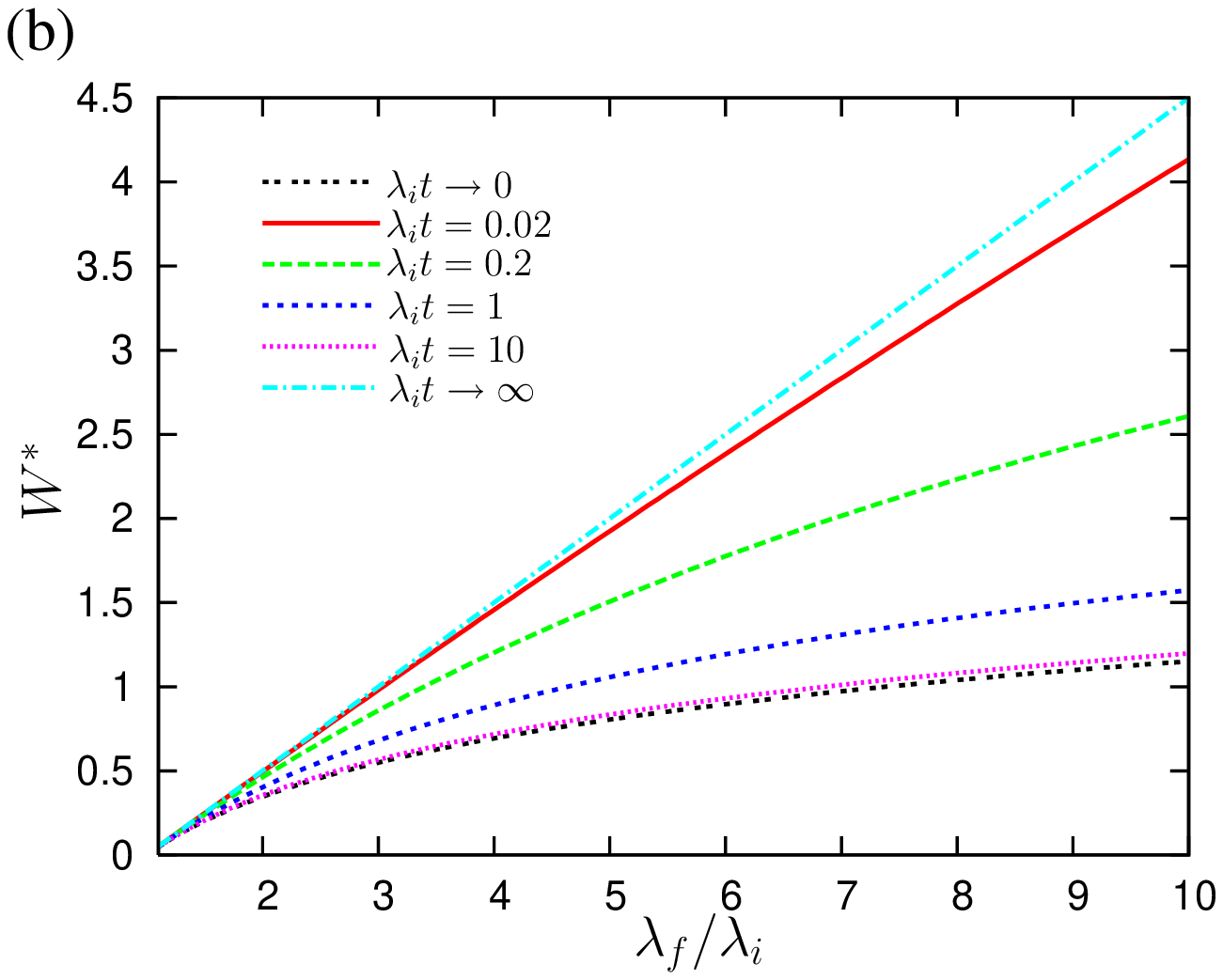}
\includegraphics[width = 0.44 \textwidth]{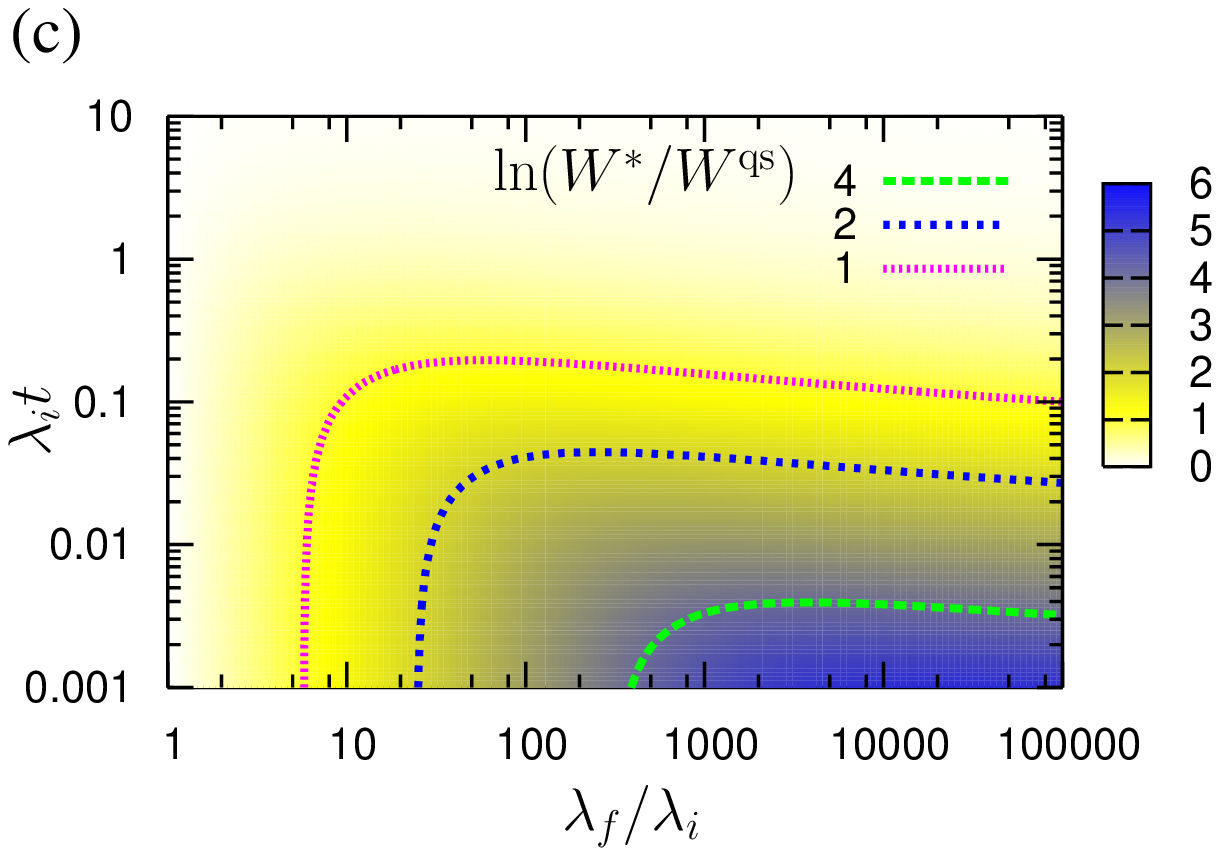}
\includegraphics[width = 0.44 \textwidth]{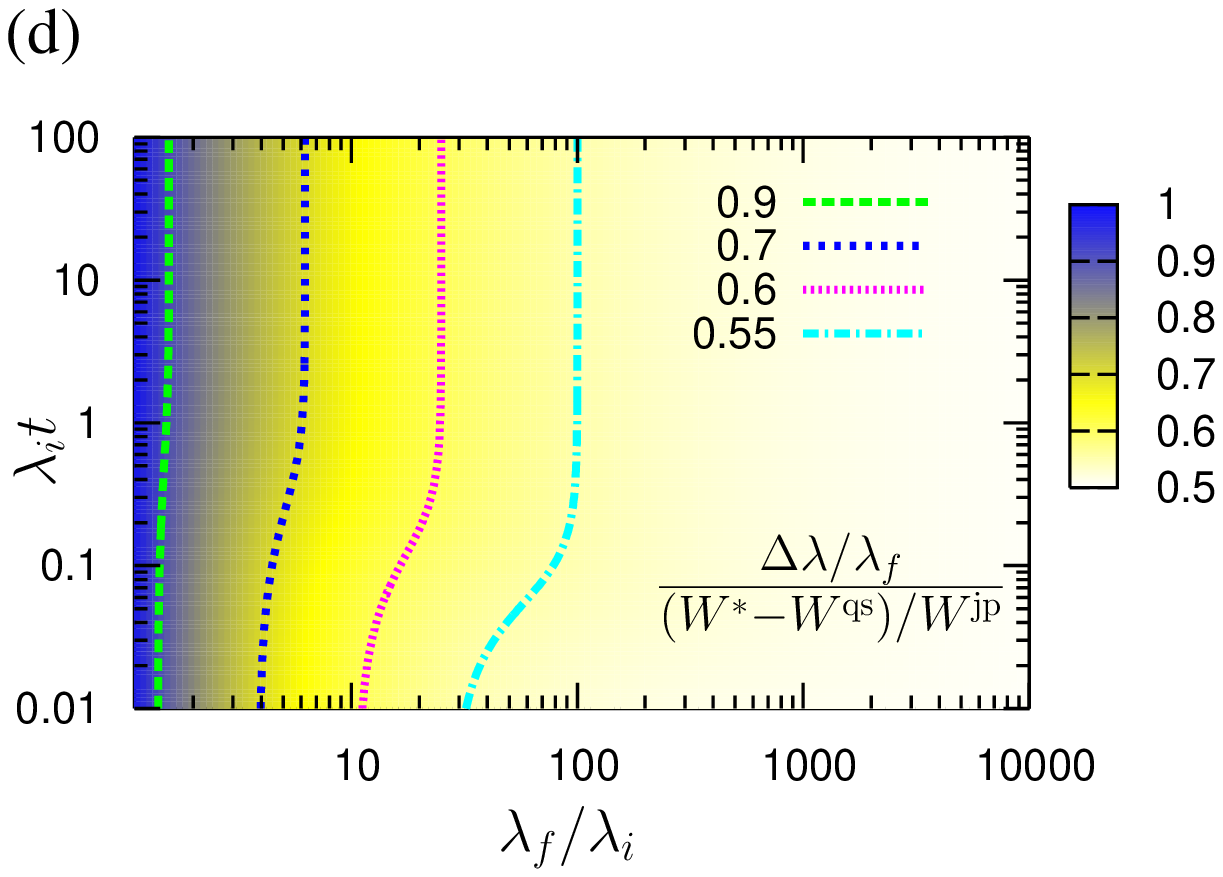}
\caption{(color online) Optimization results for a time-dependent strength $\lam(\tau)$ of a laser trap for different values of $\lf$ and  $\li t$ (case study II):
	 (a) Optimal protocols $\lam^{*} / \lam_i$ as a function of the scaled time $\tau / t$. 
	 (b) Minimal work $\Wmin$.
	 (c) Logarithmic fraction $\ln ( \Wmina / W^{\rm{\qs}} ) $ of the optimal work and the quasistatic work.
	 (d) Relative height $\Delta \lam / \lam_f$ of the jump of the optimal protocol at $t = 0$ in units of $(W^{*} - W^{\rm{\qs}}) / W^{\rm{jp}}$. 
\label{fig1}}
\end{figure*}

{\sl Case study II: Time-dependent strength of  the trap.--}
In this example, fluctuations are crucial. We consider
the motion of a colloidal particle in a trap whose strength becomes
time-dependent whereas its position remains constant. The corresponding
potential reads 
\beq
V(x,\tau) =  {\lam(\tau)} x^2 / {2} 
\eeq
with $\lam(0) = \lam_i$ and $\lam(t) = \lam_f > \lam_i$ as boundary
conditions. Such a potential with a sudden jump protocol has been investigated experimentally as test of the fluctuation theorem \cite{carb04}.
We first derive the equation of motion for the variance $\xvar(\tau) \equiv \mean{x^2(\tau)}$
\beq
\dot \xvar = -2  \lam \xvar + 2
\label{dvar}
\eeq
by multiplying eq. (\ref{FP}) with $x^2$ and integrating over $x$.
The mean work (\ref{meanw}) can then again be cast in a local functional of the new variable $\xvar$ and its first derivative by solving (\ref{dvar}) for $\dot \lam$
\begin{eqnarray}
W[\lam(\tau)] = \int_0^{t} d\tau \dot \lam \frac {\xvar} 2 
= \frac 1 2 \left [ \lam \xvar - \ln \xvar \right]_0^t + \frac 1 4 \int_0^t d\tau \frac {{\dot \xvar}^2} {\xvar}.
\label{W_wang}
\end{eqnarray}
The minimization of the work functional then requires solving the Euler-Lagrange equation
\beq
\dot \xvar ^2 - 2 \xvar \ddot \xvar = 0  .
\eeq  
Its general solution
\beq
\xvar(\tau) = c_1(1 + c_2\tau)^2
\eeq
contains two constants. 
The thermal initial distribution  $\xvar(0) = 1 / \lam_i$ fixes
$c_1=1/ \li$. The second constant $c_2$ follows from minimizing the total
mean work 
\beq
W = \frac {(c_2 t)^2} {  \lam_i t} - \ln \left (1 + c_2 t \right) + \frac 1 2 \lf  \left(1+ c_2 t \right )^2 - \frac 1 2 
\label{Wcase2}
\eeq 
which leads to 
\beq
c_2^{*} t =  \frac{-1-\lam_f t + \sqrt{1 + 2 \li t + \lam_f \li t^2}} {2 + \lam_f t} .
\label{c2}
\eeq
The optimal protocol derived from eq. (\ref{dvar})
\beq
\lmin(\tau) = \frac{ \lam_i- c_2^{*} (1 + c_2^{*} \tau)}{(1 + c_2^{*} \tau)^2}
\label{lamcase2}
\eeq
for $0<\tau<t$
again implies jumps at the beginning and end of the process as
shown in Fig. \ref{fig1}a.   
Both the minimal work $\Wmin$, see Fig. \ref{fig1}b, and the scaled optimal protocol $\lam^{*}(\tau / t) / \lam_i$ depend only on two parameters $\lf$ and $\li t$.

For the two
limiting
cases of an immediate jump, $t \to
0$, and a quasi-static process, $t \to \infty$, respectively, the values of $\Wmina$ also follow from general principles.  
For an immediate jump, the minimal work 
\beq
W^{\rm{jp}} \equiv \lim_{t\to 0} \Wmina = \mean{ \frac {(\ll_f-\li) x^2} 2 }_{\li}=\frac 1 2 (\lf -1)
\eeq
is equal to the difference in energy evaluated in the thermal initial
ensemble. 
In this limit, the optimal protocol $\lmin(\tau) / \li \approx (\lf + 1)/2$ is
constant for $0<\tau
<t$ but has discontinuities at $\tau = 0$ and $\tau = t$.
In the quasi-static  limit, the minimal work 
\beq
W^{\rm{\qs}} \equiv \lim_{t\to \infty} \Wmina = \frac 1 2 \ln \lf   = \Delta F
\eeq
is equal to the free energy difference $\Delta F$
between the final and the initial state.
In  this limit, the optimal protocol is continous at $\tau = 0$ and $\tau = t$ and takes the form
\beq
{\lmin} (\tau / t)  \approx \frac {\li}
{\left( 1-(\tau/t) +\sqrt{(\lam_i/\lam_f)} (\tau/t) \right)^2}  .
\eeq 

For $\lf \gg 1$, the minimal work is of the order of the quasi-static work for any $\lam_i t^{*} \gg 2 / \ln \lf$ as a simple analysis of eqs. (\ref{Wcase2}) and (\ref{c2}) shows. Thus, the larger the change of the control parameter $\lam$, the smaller is the time-interval required to essentially reach the quasi-static work, as quantitatively shown in Fig. \ref{fig1}c. The origin of this surprising features lies in the fact that the relaxation time scales like $1 / \lam$. For large $\lam$, the particle can follow a larger change of the control parameter almost quasi-statically. Therefore, the optimal protocol can become quite steep towards the end of the process for large $\lam_f$.

{\sl General case. --}
For a general non-harmonic potential, it is not possible to express the mean
work
as a local functional of just one variable as we have done for the two
harmonic
cases. Rather, our
optimization problem becomes non-local in time since changing the protocol at
a time $\tau$ affects the mean work increments for all later times
$\tau'>\tau$. This fact becomes obvious by expressing the mean work as
a path integral average
\beq
W[\la] = \int \dxp \pp \int_0^t d\tau \dot \lam \pd V \lam 
\label{Wm_allg}
\eeq
over all possible trajectories $x(\tau)$ with weight %\cite{chai}
\beq
\pp = \mathcal{N} p(x,0) \exp \left [-\int_0^t d\tau \left ( \frac{(\dot x +\partial_x V)^2}   { 4} - \frac{\partial_x^2 V}{2} \right ) \right ],
\eeq 
where $\mathcal{N}$ is a normalization constant. Minimizing the mean work (\ref{meanw}) then requires solving the non-local Euler-Lagrange equation 
\beq
\frac d {d \tau} \mean{\pd V \lam}_{\mid \tau = \sigma} = \frac {\delta W[\la]}
{\delta \lam(\sigma)} 
\eeq
where the right hand side can be expressed by correlation functions as
\begin{eqnarray}
&&\frac d {d \tau}  \mean{\pd V \lam}_{\mid \tau = \sigma} = ~~\dot \lam \mean{\frac {\partial^2 V}{\partial \lam^2}}_{\mid \tau = \sigma} +  \\&&\int_\sigma^t d\tau \frac{\dot \lam} 2 \mean{ \left( \frac {\partial^3 V}{\partial \lam \partial^2 x} -  \left(\dot x +  \partial_x V \right) \frac {\partial^2 V}{\partial \lam \partial x}  \right)_{\mid \tau = \sigma} \cdot {\pd V \lam}_{\mid x(\tau)}}. \nn
\label{EL_allg}
\end{eqnarray}
In general, this integro-differential equation solved by the optimal protocol
$\lmin(\tau)$ looks rather inaccessible. Exploring a variational ansatz for $\ls$ allowing for jumps with numerical evaluation of the mean work 
seems possible but may still be a formidable task to be explored in future
work.
  
In order to use jumps in the protocol $\lam(\tau)$ for the efficient extraction of free energy differences from finite-time path sampling via the Jarzynski relation, one needs an estimate for the height of these jumps without knowing the underlying potential. For the moving laser trap (case study I), we get the relation $\Delta \ll / \lam_f = 2 (W^{*} - W^{\rm{\qs}}) / W^{\rm{jp}}$. For case study II, we find numerically that the relative height of the jump $\Delta \ll / \lam_f$ at $t = 0$ is also of the order of $(W^{*} - W^{\rm{\qs}}) / W^{\rm{jp}}$, see Fig \ref{fig1}d. If such a relation gave the correct order of magnitude for the optimal jump in general cases, it could become a helpful tool for estimating the optimal jump heights. For experimentally determining the optimal protocol for an unknown potential, we envisage an adaptive procedure in which trial protocols (including estimated trial jumps) are successively improved in an iterative fashion guided by the monitored work values.

{\sl Concluding perspectives. --} 
As a main qualitative result, our analysis of two simple but experimentally
realizable model cases has revealed that the optimal protocol minimizing the mean work required to drive the system from one equilibrium state to another
involves jumps of the external control parameter both at the beginning and at the end
of the finite-time process. We expect such jumps to be a generic feature of
the optimal protocol for arbitrary potentials. Even though we have
investigated only a single degree of freedom, the extension to many coupled
degrees of freedom involves only minor notational complexity but  poses no
further conceptual challenge.

We have focussed on optimal protocols connecting two different equilibrium
states. An optimal protocol for transitions in finite-time between two
different 
non-equilibrium stationary states 
could be investigated along similar lines in the context of steady-state
thermodynamics \cite{hata01}. Likewise, one can ask for the
optimal protocol of cyclic processes 
combining mechanical steps with chemical reactions given a finite cycle time. 
These perspectives to be investigated
in future work demonstrate that the optimization problem introduced here
for stochastic thermodynamics has not only a broad fundamental significance.
Its ramifications
 could ultimately also lead to the construction of ``optimal'' nano-machines.
Finally, it is tempting to speculate which, if any,  biological processes on the
cellular and subcellular level have been optimized during evolution for their
finite-time performance in the noisy cellular environment.

\end{document}